\PassOptionsToPackage{table}{xcolor}
\documentclass[authorversion,sigconf]{acmart}

\AtBeginDocument{%
  \providecommand\BibTeX{{%
    \normalfont B\kern-0.5em{\scshape i\kern-0.25em b}\kern-0.8em\TeX}}}

\setcopyright{acmcopyright}
\copyrightyear{2024}
\acmYear{2024}
\setcopyright{acmlicensed}\acmConference[CHI '24]{Proceedings of the CHI Conference on Human Factors in Computing Systems}{May 11--16, 2024}{Honolulu, HI, USA}
\acmBooktitle{Proceedings of the CHI Conference on Human Factors in Computing Systems (CHI '24), May 11--16, 2024, Honolulu, HI, USA}
\acmDOI{10.1145/3613904.3642223}
\acmISBN{979-8-4007-0330-0/24/05}

\usepackage{xspace}
\newcommand{\ie}{i.e.{\xspace}}
\newcommand{\eg}{e.g.{\xspace}}

\newcommand{\rev}[1]{#1}
\newcommand{\qt}[1]{\textit{``#1''}}
\newcommand{\pqt}[2]{\textit{``#1''}{\,}{(#2)}}

\newcommand{\sd}[1]{$SD=#1$}

\newcommand{\iqr}[1]{$IQR=#1$}
\newcommand{\md}[1]{$MD=#1$}

\newcommand{\etal}{{\xspace}et~al.{\xspace}}

\usepackage{multirow}
\usepackage{framed,enumitem} 
\usepackage{subcaption}
\newcommand{\sysname}{Mondrian}

\begin{document}
\title{Exploring Interactive Color Palettes for Abstraction-Driven Exploratory Image Colorization}

\author{Xinyu Shi}
\affiliation{%
  \institution{University of Waterloo}
  \city{Waterloo}
  \state{ON}
  \country{Canada}
}
\email{xinyu.shi@uwaterloo.ca}

\author{Mingyu Liu}
\affiliation{%
  \institution{University of Waterloo}
  \city{Waterloo}
  \state{ON}
  \country{Canada}
}
\email{m362liu@uwaterloo.ca}

\author{Ziqi Zhou}
\affiliation{%
  \institution{University of Waterloo}
  \city{Waterloo}
  \state{ON}
  \country{Canada}
}
\email{z229zhou@uwaterloo.ca}

\author{Ali Neshati}
\affiliation{%
  \institution{Ontario Tech University}
  \city{Oshawa}
  \state{ON}
  \country{Canada}
}
\email{ali.neshati@ontariotechu.ca}

\author{Ryan Rossi}
\affiliation{%
  \institution{Adobe Research}
  \city{San Jose}
  \state{CA}
  \country{USA}
}
\email{ryrossi@adobe.com}

\author{Jian Zhao}
\affiliation{%
  \institution{University of Waterloo}
  \city{Waterloo}
  \state{ON}
  \country{Canada}
}
\email{jianzhao@uwaterloo.ca}

\renewcommand{\shortauthors}{Xinyu Shi, et al.}
\definecolor{darkslateblue}{HTML}{D01C8B}

\newcommand{\xinyu}[1]{\authorcomment{RED}{Xinyu}{#1}}
\newcommand{\jian}[1]{\authorcomment{PURPLE}{JZ}{#1}}
\newcommand{\ryan}[1]{\authorcomment{GREEN}{Ryan}{#1}}

\begin{abstract}


Color design is essential in areas such as product, graphic, and fashion design. 
However, current tools like Photoshop, with their \emph{concrete-driven} color manipulation approach, often stumble during early ideation, favoring polished end results over initial exploration. 
\rev{We introduced \sysname{} as a test-bed for \emph{abstraction-driven} approach using interactive color palettes for image colorization. }
Through a formative study with six design experts, we selected three design options for visual abstractions in color design and developed \sysname{} where \emph{humans work with abstractions and AI manages the concrete aspects}. 
\rev{We carried out a user study to understand the benefits and challenges of each abstraction format and compare the \sysname{} with Photoshop. }
A survey involving 100 participants further examined the influence of each abstraction format on color composition perceptions. 
\rev{Findings suggest that interactive visual abstractions encourage a non-linear exploration workflow and an open mindset during ideation, thus providing better creative affordance.} 

\end{abstract}

\begin{teaserfigure}
    \centering
    \includegraphics[width=.9\linewidth]{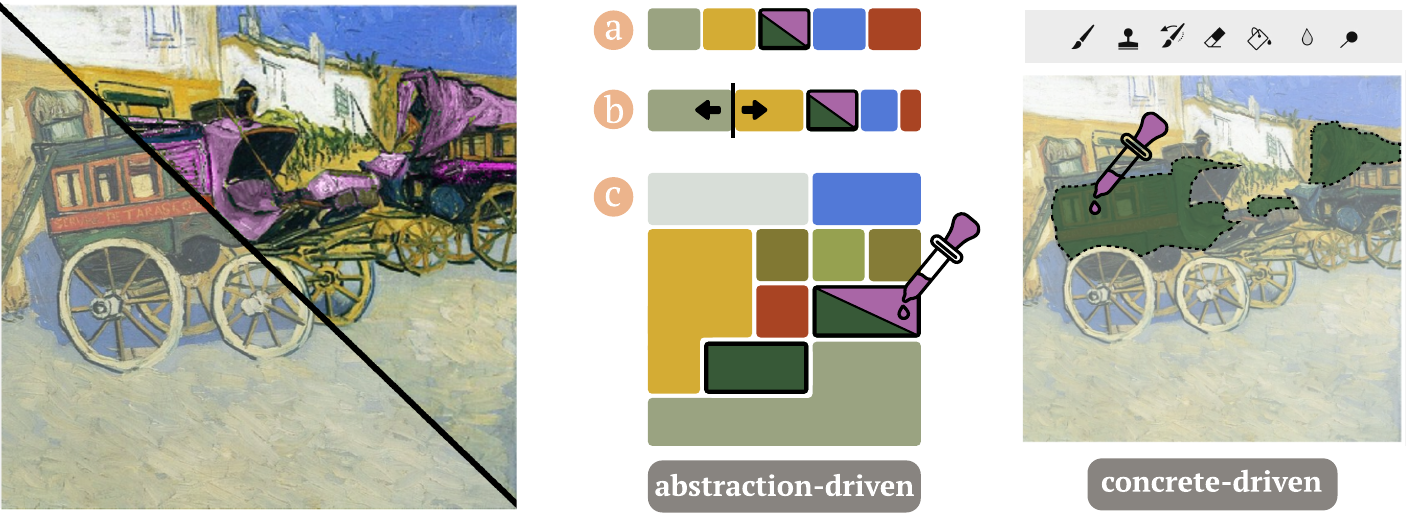}
    \vspace{-3mm}
    \caption{Demonstration of abstraction-driven and concrete-driven approach for color authoring in digital images. Users modify the interactive color palettes to recolor the image in an abstraction-driven approach while directly working on image regions in a concrete-driven approach. Three color palette formats are discussed in this paper: (a) 1D uniform (1D), (b) 1D proportional (1D+), and (c) 2D spatial (2D) palettes.
    }
    \label{fig:teaser}
    \Description{This figure illustrates two strategies for adjusting colors in digital images: abstraction-driven and concrete-driven approaches. On the left side of the figure, an image is split into two halves to illustrate the effects of color modification: the left half of this image remains in its original color scheme, serving as a baseline for comparison, while the right half has been recolored to pink demonstrate the applied changes. In the middle, the ``Abstraction-Driven Approach'' is depicted, where users adjust colors by interacting with a color palette. It is further divided into three sections, each representing a different palette format: the top section showcases a ``1D Uniform (1D)'' palette, consisting of a simple linear array of colors; the middle section displays a ``1D Proportional (1D+)'' palette, similar to the first but with color segments varying in size to represent proportion; and the bottom section presents a ``2D Spatial (2D)'' palette, which is a two-dimensional grid of colors allowing for more complex color relationships. On the right, the ``Concrete-Driven Approach'' is shown, where users modify colors by directly segmenting specific areas of the image to apply the recolorization. 
    }
\end{teaserfigure}



\maketitle

\section{Introduction}
\label{sec:intro}
Color design is essential in various fields, including \rev{graphic}, interior, fashion, and product design.
It refers to selecting and arranging colors to enhance the aesthetic appeal of a product, space, or visual content, such as graphic designs or images \cite{holtzschue2012understanding}.
As a crucial aspect of visual communication, color design can frame the work's overall mood, tone, and impact on viewers \cite{hummell2006synectics}.
Achieving a visually appealing color design often requires designers to engage in multiple iterations and experimentations.  
However, such exploration in the early ideation stage is frequently bypassed or hurriedly decided after limited attempts \cite{neeley2013building, piya2017co}, possibly due to the lack of support provided by existing tools which mostly focus on refinements for a polished final version rather than facilitating explorations.

Existing tools, for example, Photoshop, pose perceptual and operational challenges during early-stage color exploration. 
Perceptually, designers tend to base their judgment on the color combination and interdependence, such as the dominant and salient colors \cite{kim2017thoughts} and how colors interact with each other \cite{whitfield1990color}. 
However, such heterogeneous information \cite{kim2017thoughts} is not effectively reflected in existing tools; instead, they present a color summary view focusing on particular color properties (\eg, hue, saturation) which do not help the understanding of color interdependence. 
Operationally, these tools stress precision in colorization, causing designers to spend a disproportionate amount of time on overly detailed tasks (\eg, complex region selection).
Generally, these tools treat the concrete image as the primary subject of manipulation, driving users to focus on intermingled granular details such as color, shapes, pixels, edges, etc., which we referred to as a \emph{concrete-driven} approach. 
\rev{An alternative is to adopt an \emph{abstraction-driven} approach, where the users' focus shifts from a target image itself to an interactive visual abstraction representing its colors.
Visual abstraction serves a critical role in the development of ideas for encouraging divergent thinking \cite{welling2007four, root2004artistic, goldschmidt2011avoiding}.
Examples include sketch strokes in shape design \cite{pandey2023juxtaform} and diagrams in architecture planning \cite{clark2012precedents}.
Particular to the color design context, prior research \cite{shugrina2020nonlinear} demonstrates the operational benefits of using color abstraction to recolor images, thereby avoiding complex edits.
This approach, where humans interact with abstractions while AI manages the concrete aspects, offers new possibilities for supporting the early exploratory stage of color design.
However, it lacks consideration of crucial aspects such as color size and position, which have been shown critical in understanding color relationships \cite{jalal2015color}.
The current exploration of color abstraction primarily focuses on approximating the color distribution to optimize for computational precision and efficiency, rather than tailoring it to optimize for human perception of colors. 
This gap highlights the need for a more effective design of color abstractions, investigating what color aspects should be integrated to enhance human interaction and perception.
Design considerations of assessing whether a color aspect should be incorporated into the visual abstraction involve evaluating its benefits (\eg, flexibility and controllability it brings) and constraints in user interaction, as well as its impact on the human perceptual judgment of color compositions. 
These considerations remain largely unexplored in current literature.
Therefore, we are interested in four research questions to deepen our understanding and improve the practice of abstraction-driven color design: 
} 

\begin{itemize}
    \item \textbf{RQ1}: What are the potential design choices of \emph{visual abstraction} for color design?
    \item \textbf{RQ2}: What benefits and challenges does each abstraction choice present when users \emph{interact with} them? 
    \item \textbf{RQ3}: How does the \emph{abstraction-driven} manipulation influence the color design ideation process compared to \emph{concrete-driven} color editing?
    \item \textbf{RQ4}: 
    \rev{How do different abstraction choices affect human \emph{perception and understanding} of color compositions?}
\end{itemize}

To explore this paradigm and answer these questions, we began by conducting a formative study with six design experts to initially validate the feasibility of the abstraction-driven approach and identify visual abstraction design options, which resulted in the three color palette formats depicted in Figure~\ref{fig:teaser}(abc) (\textbf{RQ1}). 
Based on this, we developed \sysname{}, a system that enabled abstraction-driven image colorization by incorporating interactive color palettes and AI colorization models, through an iterative design process with the same six pilot users.  
Subsequently, we conducted an in-lab study to explore users' experiences with different visual abstraction formats (\textbf{RQ2}), and to compare abstraction-driven color manipulation enabled by \sysname{} and concrete-driven approach represented by Photoshop (\textbf{RQ3}). 
Lastly, we carried out a survey study to systematically and quantitatively assess the impact of different abstraction formats on human perception of color compositions (\textbf{RQ4}).

Our key findings reveal that working on interactive visual abstractions can better afford creative exploration by breaking the linear workflow into a more flexible, non-linear one, transitioning between the global and local perspectives.
Such a shift is also likely to lead to a receptive, exploratory mindset rather than one confined by predetermined plans.
In addition, participants highlighted that the granularity (\ie, number of colors) within each abstraction and the varied abstraction levels (\ie, different palette formats) were important. 
All individual abstraction has its benefits and challenges varying from image types and manipulation purposes. 
However, when combined as such a hierarchical abstraction framework, they not only facilitate a more profound understanding of color interplay but also enable enriched exploration.
This suggests that when designing visual abstractions, it would be necessary to identify the necessary varied abstract levels to form a spectrum of abstraction, rather than seeking an optimal abstract representation fitting for all. 

\rev{The main contribution of this paper is to provide empirical understandings of how designers explore color design choices using interactive color palettes, evidenced through a series of studies with \sysname{}.
The insights of how designers \emph{perceive} and \emph{interact with} different color abstractions highlight the need for providing a range of abstraction choices. 
This variety is crucial for enabling sufficient flexibility in exploration and offers future design directions of visual abstractions.
The comparison between \sysname{} and Photoshop illustrates how the paradigm of \emph{``humans work on abstractions while AI manages the concretes''} impacts the workflow and mindset that designers adopt during the ideation stage of color design.
The findings obtained through our studies offer implications for future research in developing abstraction-driven tools for supporting ideation in exploratory design tasks.
}



\section{Related Work}
\rev{
In this section, we motivate the design of \sysname{} by reviewing the abstraction used in creative activities, existing color palette representations, and inspirations that our tool draws from
prior research systems developed to help with color authoring. 
Furthermore, we review prior research on color aesthetics and emotional associations to contextualize our survey study on color perception.
}

\subsection{Abstraction in Creative Activities}
Abstraction \cite{palmiero2020relationships}, which reveals the essence and structure while erasing the details of things, has been considered a crucial component in creative activities \cite{welling2007four, ward2004role}. 
It benefits creative experimentation by enforcing individuals to focus on the high-level, abstract aspects of an activity, such as the purpose and objectives, rather than thinking about the low-level, concrete aspects of the task, such as the methods and techniques used \cite{root2004artistic, goldschmidt2011avoiding, dorta2008design}. 
In design and visual art, abstraction is often realized through the use of simplified forms and shapes to convey ideas or emotions \cite{ware2010visual, arnheim2004visual}, rather than providing an accurate representation of the physical world. 
For example, natural objects can be represented through a collection of simpler geometric primitives \cite{palmer1999vision} or sparse strokes \cite{pandey2023juxtaform} to help conceptual shape design.
By using abstraction, the artist or designer can focus on key elements such as color, line, and texture, rather than being limited by the constraints of realistic representation \cite{bimler2019art}.
As a result, it facilitates exploring new ideas, expressing emotions and personal experiences, and creating unique and innovative works of art and design.
Our proposed \emph{interactive abstraction for color design} is grounded by the above theories and practices, and we believe such interactive abstraction can efficiently prompt designers' creativity in color design tasks.

\subsection{Color Palette Representations}
\label{sec:palette}

Color palettes is a form of abstraction in graphic design, used to inspire designers as well as provide a convenient and organized selection of colors for a design~\cite{eckert2000sources}.
Palettes can be broadly categorized into two types: discrete and continuous.
The 1D linear format \cite{itten1970elements}, the most popular representation of a discrete palette, displays colors as constant-sized grids.
Another discrete format is the bar-graph color palette \cite{gijsenijdetermining}, which shows the proportion of colors using varied lengths of bars, with each color corresponding to one bar and its length indicating the percentage of that color w.r.t. the whole palette.
The 2D color palette, proposed by Shi~\etal~\cite{destijl2023shi}, incorporates the proportion and the spatial placement of colors into the palette representation. 
Different from the discrete ones widely applying for the general design scenarios, continuous palette frequently designed for a specific domain. 
The color ramp \cite{smart2019color}, presents a spectrum of colors sorted by gradients, commonly used to encode quantitative data in visualizations.
Color Triads \cite{shugrina2020nonlinear} employs a triangular-shaped palette to depict non-linear blending of colors for shaded objects. 
While it is continuous, its format is sparse and structured. 
In contrast, the Playful Palette \cite{shugrina2017playful} offers a dense representation, capturing colors in artwork in a freeform and unstructured continuous format.

Although various color palette formats have been proposed, there is a lack of comprehensive studies investigating their impact on users' perceptions, workflows, and behaviors, an area we aim to address in this paper.
In specific, we aim to investigate the role of these different formats of color palettes as abstractions of images to aid in color design and how they influence users to perceive and interact with different color compositions.


\subsection{Interactive Tools for Color Design}
Color design (\ie, involves picking the colors and then colorizing the image) plays a crucial role in a designer's workflow, as it greatly influences visual communication and aesthetics. 
To create harmonious and engaging color compositions, designers often rely on color palettes as a foundation. 
Color picker interfaces \cite{olsen2007evaluating} have evolved over the years to allow for individual color sampling. 
Some advanced interfaces have been designed to support color tweaking \cite{shugrina2019color} and mixing \cite{shugrina2017playful}, particularly for applications like vector graphics coloring and artwork drawing.
However, designers tend to evaluate colors within a composition as a whole rather than as individual colors \cite{whitfield1990color}, since the interplay between colors can significantly impact the overall aesthetic and emotional impact of the composition \cite{eiseman2000pantone}. 
The arrangement, balance, and combination of colors can evoke diverse emotions, communicate distinct messages, and guide the audience's gaze \cite{frascara2004communication}. 
Consequently, considering colors within the context of the entire composition is crucial for informing design decisions \cite{barnard2013graphic}. 

To facilitate designers with making decisions, Interactive Palette Tools (IPTs) \cite{meier2004interactive} was developed two decades ago to support the exploration of color composition, mixing, and palette assignment with separate tools.
Similarly, Color Portraits \cite{jalal2015color} characterizes the design space for five key color manipulation activities and emphasizes the importance of \emph{in-context exploration}, allowing users to manipulate the relationships of colors through interactive palettes.
However, the ``context'' here only includes the color fields but not the corresponding images, resulting in a fragmented workflow for color design---the color composition exploration and the adaption of the palette to the target design need to be done separately.
Color Triads \cite{shugrina2020nonlinear} enables the manipulation of color distribution on a triangular-shaped palette and automatically colorization for the target design. 
Histomages \cite{chevalier2012histomages} allows for recoloring images through interactive color histograms. 
These tools have presented a unified color design workflow; however, neither the color triads representation nor the color histogram can intuitively reflect how humans perceive colors in an image. 
For example, the information on size, shape, and placement of colored objects in an image is missing from these two representations, which are revealed in the 2D palette mentioned earlier.

In this paper, we explore what color palette formats, as an image abstraction, are intuitive for users to experiment with color compositions, and how such abstraction can be seamlessly integrated with the image colorization process through user-driven exploration to support a unified color design workflow. 

\subsection{Color Aesthetic and Emotions}
The selection of colors for aesthetic appeal can be guided by either rule-based color harmony theories or data-driven computational models \cite{holtzschue2012understanding}. 
Firstly, rule-based theories \cite{wyszecki2000color}, such as complementary, split complementary, triadic, tetradic, and analogous color harmonies, are based on the relationship between colors on the color wheel and their relative position \cite{parkhurst1982invented}. 
For instance, complementary color harmonies use two colors that are opposite each other on the color wheel, while analogous color harmonies use colors that are adjacent to each other. 
Besides the positions on the color wheel, other color features, such as color distances \cite{moon1944geometric} and balance \cite{burchett1991color}, are introduced to heuristically determine color harmony. 
In contrast, data-driven approaches \cite{lu2015discovering, nishiyama2011aesthetic} rely on large datasets to learn the implicit compatibility patterns according to image statistics.
An example is the work by O'Donovan~\etal~\cite{o2011color}, which establishes scoring models for color compatibility measurement. 

In addition to aesthetics, colors are often used to convey emotions.
Research has shown that different color categories can evoke different emotional feelings \cite{von1970theory}, with certain colors (such as red and yellow) eliciting systematic physiological reactions resulting in emotional experience (\eg, negative arousal), cognitive orientation (\eg, outward focus), and overt action (\eg, forceful behavior) \cite{goldstein1942some}. 
Similarly, color combinations can have an impact on affective impressions \cite{bartram2017affective} when applied to visualizations, with complementary colors (\eg, red and green) creating a dynamic and energetic impression, while monochromatic color schemes (using only variations of a single color) can create a peaceful and harmonious impression.
Additionally, cultural and personal \cite{king2005human} experiences can also play a role in the emotions that colors and color compositions evoke, making the impact of color subjective and complex.

These studies contribute valuable knowledge to the field of color design; however, they do not specifically explore the influence of color palette formats, particularly the proportions and spatial placements of colors. 
This gap inspired us to investigate whether presenting color compositions in different formats can influence human perceptions of aesthetics and emotion associations, ultimately enhancing our understanding of the various factors that impact color design decisions.

\section{Overview of Methodology}
We investigate four research questions in this paper (stated in Section~\ref{sec:intro}).
To explore reasonable visual abstraction formats serving as interactive palettes (\textbf{RQ1}), we first conducted a formative study with 6 participants with different expertise levels in image color design.
We summarized the three main insights derived from the interviews and guided the development of \sysname{} to further test the abstraction-driven approach.
Further, we conducted a user study with 12 participants consisting of two tasks to answer the RQ2 and RQ3. 
Specifically, the first task in the user study aims to further compare the different formats of visual abstractions
(\textbf{RQ2}), and the second task compared \sysname{} with Photoshop which serves as a typical and popular concrete-driven approach baseline to understand how abstraction-driven manipulation may impact the color design ideation process (\textbf{RQ3}). 
Lastly, we analyzed the perceptual impacts of different abstraction formats through a survey study with 100 participants (\textbf{RQ4}).

\section{Formative Study}

\begin{figure*}[htb]
    \includegraphics[width=\linewidth]{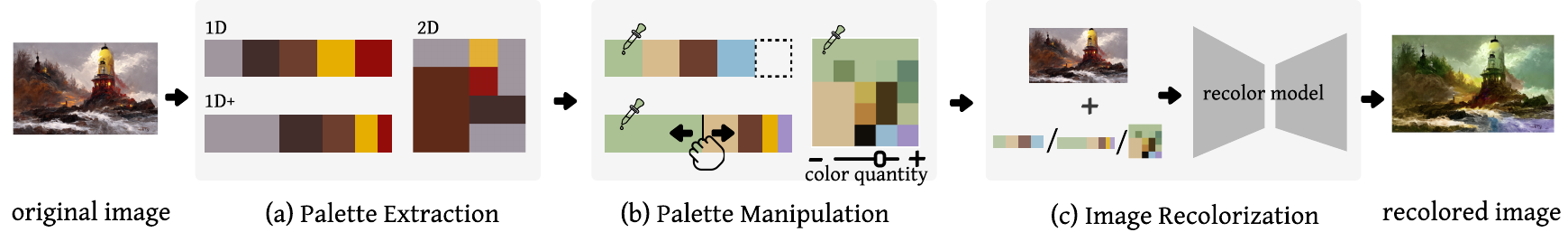}
    \vspace{-3mm}
    \caption{\rev{The system pipeline of \sysname{}, showing the main modules: (a) Palette Extraction which extracts 1D, 1D+, and 2D palettes; (b) Palette Manipulation, where users can modify colors and adjust quantities and proportions (specific to 1D+ palettes); and (c) Image Recolorization, where the recolorization model uses the modified palette and original image to generate the recolored image.}}
    \label{fig:overview}
    \Description{This figure presents the workflow of the \sysname{}, organized into three main modules. Starting from the left, there is an original image followed by the first module ``Palette Extraction'', where the system extracts color palettes in three formats: 1D, 1D+, and 2D. Next, in the center, is the ``Palette Manipulation'' module, where users can alter colors through the palettes. Finally, on the right, the ``Image Recolorization'' module takes the modified palette and the original image to produce the recolored image, showcasing the result of the process. Each module is sequentially connected, illustrating the flow from palette extraction through manipulation to the final recolorization.}
\end{figure*}

A formative study was conducted to gather insights from experts and novices in the field of image colorization, with the aim of 
seeking their feedback on the proposed \emph{abstraction-driven} approach and their thoughts on the possible visual abstraction design choices (\textbf{RQ1}).

\subsection{Participants and Procedure}

Semi-structured interviews were conducted with six participants (5 females and 1 male) consisting of graphic designers and photographers experienced in image colorization.
Two participants were novices with a keen interest in image editing and photography, while two others were expert designers boasting over five years of experience in graphic design. 
The remaining two participants fell between the beginner and intermediate levels, each with approximately one year of design experience.

The interviews began with background questions about participants' professional roles and experiences in image editing. 
After that, we introduced the concept of \emph{abstraction-driven} color editing approach, both verbally and using a low-fidelity Figma prototype. 
Participants were encouraged to share their perspectives on this.
Moreover, they were asked about the potential of utilizing color palette formats discussed in Section~\ref{sec:palette} as visual abstractions \rev{by showing them the continuous color ramp generated by Adobe Color using the ``Extract Gradient'' feature\footnote{\url{https://color.adobe.com/create/image-gradient}} and the Color Triad \cite{shugrina2020nonlinear} through the online demo\footnote{\url{https://www.colorsandbox.com/recolor}}}. 
Participants were compensated CAD\$ 30 for their time and effort.
Our findings have been distilled into valuable insights to guide visual abstraction design contributed by participants.

\subsection{Design Insights}

Three key design insights were contributed by participants, which guided the development of the final \sysname{} system. 

\begin{itemize}[leftmargin=2em]
    \item[\textbf{D1.}] \textbf{Use discrete color palettes as visual abstractions.}
    Both discrete and continuous palettes (described in Section~\ref{sec:palette}) offer advantages and drawbacks, but discrete palettes are particularly suited for image colorization. 
    Their straightforward structure is easy to extract from images and offers designers a clear insight into color composition. 
    As noted by E2, \qt{I usually pay more attention to colors with a large portion of high-contrasted colors attracting the focus.}
    Their adaptable nature makes interacting and adjusting color numbers simple. 
    Conversely, continuous palettes, like two-color ramps or three-color triads, have a base color count, largely limiting their adaptability for complex images which often have more varied colors.
    Further, designers also demand to \pqt{quickly locate a certain color [in the image] of interest on the palette, but hard to do so on it (the continuous palette)}{E4}.
    We thereby primarily focus on discrete palettes as visual abstractions.

    \item[\textbf{D2.}] \textbf{Offer various palette formats for versatile design exploration.}
    Among the discrete palettes: 1D \cite{itten1970elements}, 1D+ \cite{gijsenijdetermining}, and 2D \cite{destijl2023shi}(Figure~\ref{fig:teaser}(abc)), their feedback was enlightening: instead of viewing one palette as superior to the others, they perceived each format as possessing its own set of strengths tailored for particular design situations. 
    For instance, \pqt{the 1D might be suitable for simplistic designs}{E1}, whereas \pqt{the 2D offers a broader canvas for intricate projects requiring multiple color interplays}{E3}.
    Thus, they suggested to integrate the three formats, rather than sticking to a single one, would be a promising approach to offer comprehensive control over various design perspectives.

    \item[\textbf{D3.}] \textbf{Provide controls at different granularities of abstraction.}
    In our initial prototype of \sysname{}, we provided five colors for each color palette format, which was inspired by traditional platforms like Adobe Color. 
    However, by interacting with our users, we found that a more diverse range of granularity was needed for the interactive color palettes. 
    This is because the complexity of color schemes in different images varies, and thus the number of colors required for effective control may differ. 
    Additionally, different design stages, such as initial exploration and fine-tuning, demand varying levels of control granularity. 
    We thus refined our system to offer adjustable color palette sizes, enabling users to have a customizable and flexible level of control over their unique design needs.


\end{itemize}

\section{\sysname{}}



In this section, we present the detailed design and development of the \sysname{} system. 
We provide a high-level system overview, demonstrate the interface through a usage scenario, and explain the backend techniques, including palette extraction and image recolorization processes.

\subsection{System Overview}
\begin{figure*}
    \centering
    \includegraphics[width=1.\linewidth]{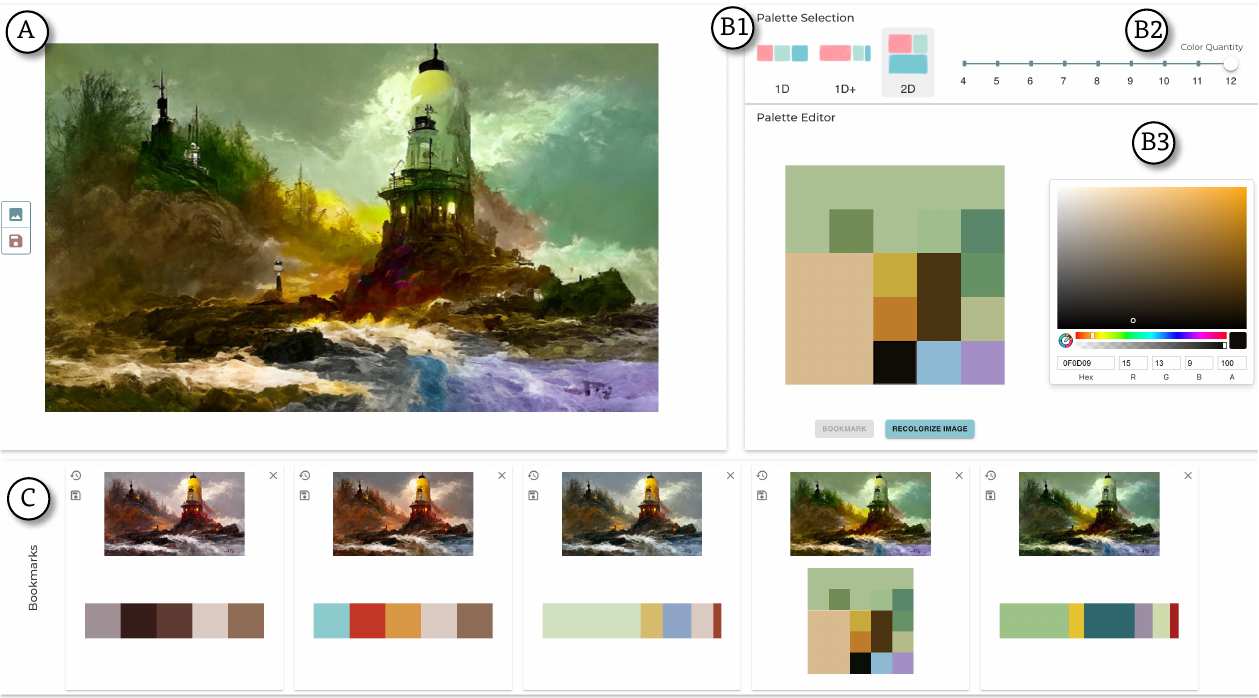}
    \vspace{-6mm}
    \caption{The user interface of \sysname{}, consisting of Result Preview (A), Palette Manipulation (B), and Bookmark Tracking (C) panels. Palette Manipulation offers interaction with three palette formats (B1): 1D uniform (1D), 1D proportional (1D+), and 2D spatial (2D) palettes, adjustable with a global slider (B2). The Bookmark Tracking (C) panel stores intermediate states for assisting creativity exploration.
    }
    \label{fig:interface}
    \Description{The figure illustrates the user interface of \sysname{}, organized into three key panels. At the top-left is the ``Result Preview'' (Panel A), displaying the outcome of color adjustments. Next to it, on the right, is the ``Palette Manipulation'' section (Panel B), which is further divided into three areas: the first area (B1) presents options for interacting with three types of color palettes - 1D uniform (1D), 1D proportional (1D+), and 2D spatial (2D); the second area (B2) features a global slider for adjusting the color quantity in the palettes; and below them, is the third area (B3): palette editor, where users directly manipulate on the palettes. At the bottom is the ``Bookmark Tracking'' (Panel C), a space used for saving and organizing intermediate states to facilitate the exploration of creative possibilities.}
\end{figure*}

Based on the derived design rationales, we propose \sysname{}, a system that augments color palettes with interactivity to facilitate the exploration of digital image colorization possibilities. 
The front-end interface of \sysname{} comprises three panels: (A) Result Preview, (B) Palette Manipulation, and (C) Bookmark Tracking, as illustrated in Figure~\ref{fig:interface}.
The backend consists of two primary modules: palette extraction and image colorization.

We demonstrate the three key modules in \sysname{} in Figure~\ref{fig:overview}: palette extraction, palette manipulation, and image recolorization.
Upon uploading an image, the palette extraction module processes it to generate 1D, 1D+, and 2D palettes, the approach will be explained in later sections.
These palettes are then displayed in the Palette Manipulation panel in the interface, where users can edit individual colors and, for 1D+ palettes, adjust color proportions. 
Once users create a satisfactory palette, it is passed along with the image to the image recolorization module, which recolors the image to match the given palette. 
The recolored image and its corresponding palette are stored in the Bookmark Tracking panel for easy access and comparison.

\subsection{User Interface}

In the following, we walk through the \sysname{} interface with a simple usage scenario.


Suppose a graphic designer, Alice, who wants to recolor an image for a new project. 
She first uploads the image using the Result Preview panel (Figure~\ref{fig:teaser}-A), resulting in a display of the image with \sysname{}. 
Alice then starts to interact with the Palette Manipulation panel (Figure~\ref{fig:teaser}-B), where she explores the three available color palette formats: 1D uniform (1D), 1D proportional (1D+), and 2D spatial (2D). 
Using the ``1D'' option, she changes the brown to blue, red, and orange to introduce a more colorful effect. 
By clicking the ``RECOLORIZE'' button, the image is automatically recolored and shown on the Result Preview panel (A).
The image has been enhanced with added red and orange tones while the sea takes on a blue hue, evoking the feeling of a warm, inviting sunset.
She appreciates the current result but wants to explore a different ambiance with a cooler and calmer atmosphere.
She transits to the ``1D+'' palette and adjusts the color quantity using the global slider (Figure~\ref{fig:teaser}-B2) to 7 for a more intricate control over the color distribution.
She specifies a new group of colors and increases the proportion of green by dragging the edges of the color block, which makes the image dominant by green more. 
The resulting image transforms the hill from orange to green, creating an overall tone that evokes a more tranquil feeling.
However, she feels the current grey sky is quite boring.
Thus, she experiments with the ``2D'' palette, which allows her to manipulate color placement for tuning colors for local regions in the image.

As she iterates on her design, she stores various intermediate states as bookmarks using the Bookmark Tracking panel (Figure~\ref{fig:teaser}-C). 
This enables her to easily compare different design variations and returns to any previous state for further modifications.
She frequently switches between the three palette formats to make adjustments and find inspiration, with each format offering a unique perspective on color composition including aesthetics and affectiveness. 
This interplay between the palettes helps Alice discover serendipitous color combinations and encourages further tweaking of her design.
As a result, she produces several distinct variations to present and discuss with her clients, where the overall creative process is easy and effective.

\subsection{Palette Extraction} \label{sec:palette_extraction}
In this section, we describe how the color palette in each format (\ie, 1D, 1D+, and 2D) is generated from an image. 

\textbf{1D Uniform Palette.}
The 1D palette is a collection of the top-$N$ dominant colors in an image arranged in equal-sized blocks. 
To extract the dominant colors, we first convert the image to the $Lab$ color space and apply k-means clustering to group the pixels into $N$ clusters. 
We then determine the frequency of colors within each cluster and select the most frequently appeared color as the dominant color of that cluster. 
These dominant colors are arranged in equal-sized blocks to form the 1D uniform color palette.

\textbf{1D Proportional Palette.}
The approach for extracting the 1D+ palette is similar to the above. 
The main difference is that the corresponding proportions (\ie, the relative frequency) of each dominant color are calculated and reflected in the size of each color block. 
Therefore, the 1D+ palette provides a more informative representation of color distribution in an image. 

\textbf{2D Spatial Palette.}
To extract a 2D palette from an image, we adopt the approach proposed by Shi~\etal~\cite{destijl2023shi}. 
First, we simplify the image by dividing it into $N$ superpixels using the SLIC algorithm \cite{achanta2010slic}. 
The pixels are clustered in a combined five-dimensional space consisting of the color channels $(l, a, b)$ and the pixel positions $(x, y)$, and each superpixel's dominant color is computed using kNN clustering \cite{fix1989discriminatory}. 
After downsampling to a size of $5\times5$ for abstraction and upsampling back to the original size, semantic regions are transformed into regular-shaped color blocks, resulting in the 2D palette.
The 2D spatial palette incorporates spatial placement information of colors, aligning the layout of color blocks with the image structure. 

\subsection{Image Recolorization}
To automatically apply user-edited palettes to images, we employ off-the-shelf data-driven colorization models for image recolorization based on given palettes. 
Specifically, we use HistoGAN \cite{afifi2021histogan} to recolorize images with 1D and 1D+ palettes with the pre-trained checkpoint of ``Universal model-0'', and Zhang~\etal's method~\cite{zhang2017real} for 2D spatial palettes.
Both methods are based on generative adversarial networks (GANs) but focus on optimizing different aspects of color distribution. 
The 1D+ palette necessitates that the recolored image matches the specified color proportions, while the 2D palette requires alignment with the given spatial constraints.
HistoGAN achieves this optimization by projecting color histogram features into the model's latent space and designing a loss function that explicitly enforces the recolored image to align with the target color proportions. 
In contrast, Zhang~\etal's approach optimizes for color position alignments by developing an end-to-end framework that propagates sparse user-specified color points from large-scale data to predict the mapping from grayscale images to full-color images directly.

\begin{figure*}[tb!]
    \centering
    \includegraphics[width=\linewidth]{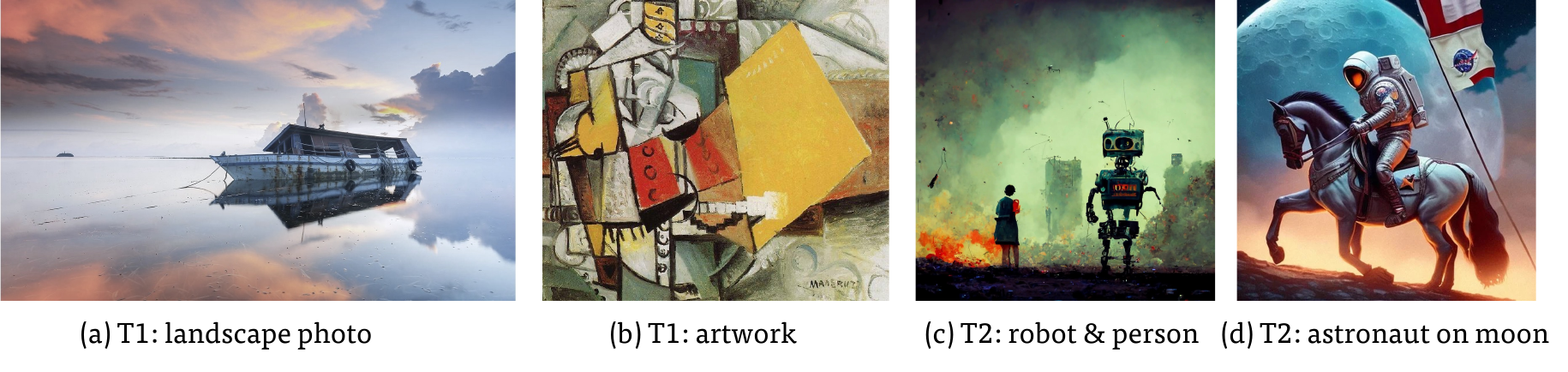}
    \vspace{-8mm}
    \caption{The images used in the user study, where (a) landscape photo and (b) artwork are for Task 1, and (c) robot with a person and (d) astronaut on the moon are used for Task 2.} 
    \label{fig:original_image}
    \Description{The figure displays four images utilized in a user study, arranged to correspond with two tasks. For Task 1, the images include (a) a landscape photo showcasing natural scenery, positioned first, and (b) an artwork featuring an abstract style painting, placed second. For Task 2, the images are (c) a robot interacting with a person, shown third, and (d) an astronaut standing on the moon's surface, displayed last.}
\end{figure*}

\section{User Study}

We conducted an in-lab user study consisting of two tasks to further understand how abstraction-driven manipulation facilitates the color design process and compare it with Photoshop. 
We chose Photoshop as it is the most popular industry-level tool for implementing the concrete-driven approach.
It should be noted that the aim of comparing with Photoshop is not to prove one system superior to the other but to discern the impacts of the two interaction paradigms (abstraction-driven and concrete-driven) on user behaviors and mindsets. 

\subsection{Participants and Apparatus}
We recruited 12 participants (\rev{6 females, 6 males}) aged 22 to 39, with a mean age of 26.6 years. 
We assessed their design experience based on a pre-study demographics questionnaire. 
It shows that 50\% of them had prior experience or had studied in an art or design-related field.
To evaluate the participant's level of experience in design, we used a 5-point Likert Item, where 1 represents ``No experience'' and 5 represents ``Expert''.
Based on their self-reported results, most participants were at a beginner or intermediate level (\md{2}, \iqr{1.5}).
All participants had attempted digital image colorization in the past six months. 
The most commonly used tools for digital image colorization among the participants were Adobe Photoshop, GIMP, Adobe Lightroom, and built-in photo editing apps on smartphones.
We refer to participants by `P' suffixed by the participant number.
Participants could access the \sysname{} through an internet browser by entering a specific web URL. 
For those who already had Photoshop installed on their local devices, we conducted an online study. 
For participants who did not have Photoshop installed, we carried out an in-person study, providing access to the software. 


\subsection{Tasks and Design}
To answer the RQ2 and RQ3, we designed two image colorization tasks correspondingly in our study. 
Images used during the tasks are shown in Figure~\ref{fig:original_image}.



\textbf{Task 1:} 
One main goal of this task is to investigate how participants use the three interactive color palette formats to complete image colorization.
Another goal was to familiarize participants with \sysname{} and allow them to explore the characteristics of the three palette formats, which prepared them for Task 2. 
Each participant was given an image to colorize with a desired style guidance, and they had to use each color palette format to generate one design (\ie, in total, three designs).
The order of palette formats being used was counterbalanced across participants.
Since the image type may potentially influence users' experience with different palette formats, we prepared two different types of images, realistic landscape photography, and abstract artwork, as shown in Figure~\ref{fig:original_image}.
The landscape photo requires participants to recolor a more visually striking one to evoke stronger emotions while the artwork image requires them to shift the color palette to express an entirely new artistic theme or concept.
We adopted a within-subjects design for this image type factor.  
Participants were not allowed to use the three palette formats in combination (while \sysname{} supports it), because we wanted to independently explore the strengths and weaknesses of each format.  
Participants had up to 2 minutes for each color palette format. 

\textbf{Task 2:} 
This task is designed to understand the pros and cons of our proposed abstract manipulation and the traditional direct manipulation in color design.
We thus compared users' experiences and behaviors in \sysname{} and Photoshop using a within-subjects design.
We simulated a design scenario where a graphic designer is tasked with modifying AI-generated images for a robot-making start-up's promotional campaign. 
The designer must enhance the color palette of two images, one featuring a robot-human interaction in a futuristic setting and the other depicting a robot astronaut riding a horse on the moon, to create eye-catching visuals and convey an inspiring spirit (Figure~\ref{fig:original_image}).
They were asked to explore possible color design solutions for 2-5 alternatives to prepare for the discussion with the client.
Participants need to deal with one image in \sysname{} and another with Photoshop.
The order of using the two systems was counterbalanced across participants.
Participants could use the three palette formats freely (\eg, in combination or not) to generate their designs. 
This allows us to explore the full range of functionality and flexibility that \sysname{} provides in coloring images with different palette formats.
Participants were given 10 minutes to complete the design for each system.

\begin{figure*}[htb!]
    \centering
    \begin{subfigure}[c]{0.55\linewidth}
        \includegraphics[width=\linewidth]{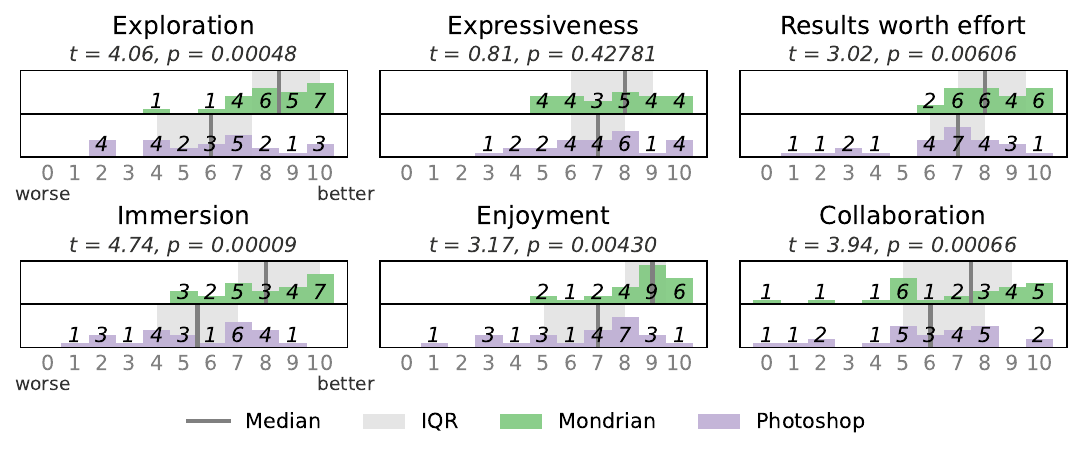}
        \vspace{-7mm}
        \caption{} \label{fig:CSI_subfields}
    \end{subfigure}
    \begin{subfigure}[c]{0.4\linewidth}
        \includegraphics[width=\linewidth]{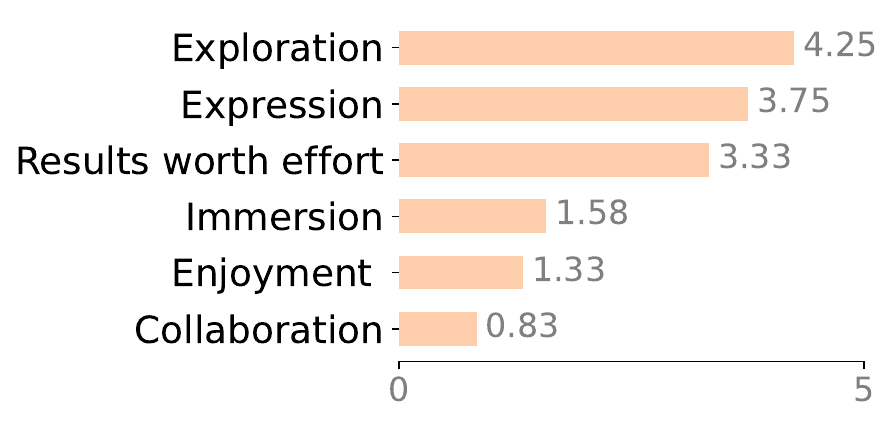}
        \caption{}  \label{fig:CSI_weights}
    \end{subfigure}
    \vspace{-4mm}
    \caption{(a) Results of Creative Support Index (CSI) for \sysname{} and Photoshop (the higher the better) for the factors of exploration, expressiveness, results worth effort, immersion, immersion, enjoyment, and collaboration. (b) Participant-rated weights of each factor of CSI.} 
    \Description{The figure is divided into two parts, illustrating the evaluation of the Creative Support Index (CSI) for the \sysname{} in comparison to Photoshop. Part (a) presents the CSI results for both systems, highlighting seven factors: exploration, expressiveness, results worth the effort, immersion, enjoyment, and collaboration. The values are plotted on a graph, where a higher score indicates better performance. Part (b) shows a graph of participant-rated weights for each CSI factor, indicating the importance participants assign to each factor in their creative process. 
    }
\end{figure*} 

\subsection{Procedure}
After signing the consent form, participants were given an introduction to the study procedure, duration, data collection, and other important information. 
Participants were then provided with a brief training session on both \sysname{} and Photoshop, during which an example was used to demonstrate how to colorize an image with each tool. 
Participants were allowed to play with the tools using pre-determined design examples independent of the user task images. 
Once they became comfortable using the tools, participants were introduced to the above tasks sequentially. 
During both task sessions, participants were asked to ``think aloud.''
After completing the two tasks, participants were asked to rate their experiences with \sysname{} and Photoshop using a post-study questionnaire, which included the Creativity Support Index \cite{cherry2014quantifying} rating on Likert scales. 
To gain a deeper understanding, semi-structured interviews were conducted to collect their feedback on both systems. 
The study lasted about 90 minutes, and all study sessions were audio and video recorded. 
Participants were compensated with \rev{CAD\$ 30} for their time.

\subsection{Quantitative Results}



In the following, we report participants' ratings of their experience with the two study systems using the Creativity Support Index (CSI).   
These quantitative results are part of the answers for \textbf{RQ3}.

\textbf{Creativity Support Index.}
The participants were asked to rate six different factors according to CSI, namely enjoyment, exploration, expressiveness, immersion, collaboration, and results-worth-effort, on a Likert scale from 0 (strongly disagree) to 10 (strongly agree). 
The results of each factor are shown in Figure~\ref{fig:CSI_subfields}. 
In addition, participants completed a paired-factor comparison section where each factor was paired against every other factor for a total of 15 comparisons, as shown in Figure~\ref{fig:CSI_weights}. 
This allowed them to weigh the factors in the context of the specific colorization task they were performing.
We followed the instructions in \cite{cherry2014quantifying} to obtain the final CSI score using the above information.


Overall, our system provided significantly greater creativity support, with a final CSI score of 83.08 (\sd{12.18}) for \sysname{} compared to 64.06 (\sd{20.38}) for Photoshop. 
The paired t-test showed a statistically significant difference ($t = 2.88, p = 0.015$) on overall CSI scores between the two systems, suggesting that \sysname{} may be a more effective tool for fostering creativity.
When examining the individual factors, we found that \sysname{} significantly outperformed Photoshop in the factors of exploration, enjoyment, results worth effort, immersion, and collaboration. 


\rev{
\sysname{} significantly outperformed Photoshop in Exploration, likely attributed to its exploratory palettes and straightforward interface.
The design of \sysname{} prioritizes simplicity and rapid experimentation over detailed control, making it more adapted to quick exploratory tasks.
In addition, participants engaged more in color experimentation with \sysname{} but tended to be more easily distracted from the tedious operations in Photoshop, resulting in a significant difference in the Immersion.
\sysname{} also scored higher than Photoshop on the factor of Results Worth Effort. 
A key reason for this might be because of Photoshop's steeper learning curve. 
Given the limited duration of the study, participants might find it challenging to fully utilize Photoshop's complex features, impacting its scores negatively on this aspect. 
We also observed greater performance on Enjoyment with \sysname{}, potentially due to its user-friendly design that may reduce participants' frustration on figuring out different functionalities in Photoshop, thus enhancing the overall experience.
Regarding Collaboration, despite the lack of direct collaborative tasks in the study, \sysname{} still received a higher score. 
This could be because its simplicity and quick exploration features were perceived as advantageous for team brainstorming and idea sharing. 
However, it is important to note that further investigation is needed in a collaborative setting to validate these findings comprehensively.

Despite these differences, we did not find a statistically significant difference between the two systems in terms of Expressiveness. 
This might be because Photoshop incorporates extensive color adjustment features for the most fine-grained control over the image, and this level of detail in manipulation allows for a high degree of expressiveness. 
Conversely, while \sysname{} offers a palette-based solution with varying levels of abstraction, it does not ensure that the detailed effects can perfectly align with users' intended outcomes, thus rated lower in expressiveness.
This observation indicates the inherent trade-off between expressiveness and exploration in design tools. 
Enhanced expressiveness, or the degree of control, often requires more complex features, which in turn can increase the effort required for manipulation. 
Such complexity may compromise exploration, which benefits from simplicity, ease, and speed of use.
We further discuss this aspect in \autoref{sec:design-considerations}.
}

\subsection{Qualitative Feedback}

We employed a thematic analysis approach to examine the participants' responses. 
Two researchers independently coded the data, identifying recurring themes and patterns in the feedback. 
The researchers then discussed and reconciled any discrepancies in the coding process to ensure a consistent and accurate representation of the participants' perspectives. 
\rev{
In the following, we analyze the differences among the three color palettes derived from Task 1 to answer RQ2, with detailed discussions presented in Section~\ref{sec:finding_task1}. 
Subsequently, to answer RQ3, we elaborate the comparison between \sysname{} and Photoshop in the context of Task 2 in Section~\ref{sec:finding_task2}.
}

\subsubsection{User Interactions across Different Palette Formats (\textbf{RQ2})}
\label{sec:finding_task1}
Based on our observation and participants' comments, we identified the following key aspects of the differences in user interactions across the three interactive color palette formats. 

\textbf{Intuitivieness.} 
In general, between two 1D formats, participants consistently favored the proportional 1D palette over the uniform one, as \pqt{it more effectively revealed the mapping between the palette and the image and intuitively presented the color distributions}{P1}.
All participants agreed that the spatial 2D format provided the most intuitive experience due to its precise mapping of color placements. 
However, we observed two trends of user behavior patterns, leading to varying opinions regarding the 1D proportional format. 

Some participants (4/12) tended to seek a more controlled and targeted way of colorization, focusing on individual regions and their specific colors. 
This group found the proportional 1D palette a bit confusing at first, as P7 stated that \qt{initially, I wondered how to map each color block to a specific region in the image; thus, I made changes very carefully to only one block at a time to see where is the exact corresponding region.} 
In contrast, these participants found the 2D format presents \pqt{a more direct connection between the palette and the image}{P5}, and allows them \pqt{to precisely understand which regions of the image would be affected by my manipulation}{P7}.

On the other hand, other participants preferred a more holistic view of the image, with less emphasis on precise color assignments for individual regions, finding the 1D proportional palette \qt{straightforward} and \pqt{perfectly compliments with 2D one for different modification purposes}{P3}.
They opted first to modify all colors simultaneously to specify an overall color combination using the 1D proportional palette. 
\pqt{I felt the fuzzy mapping between the 1D [proportional] format and the image is good enough for sensing and adjusting the overall tone}{P6}.
After getting a satisfactory overall tone, they would refine some local parts using the 2D format and felt \pqt{easy to identify small regions I want to add more contrast to make something stand out on the 2D palette}{P4}.

\begin{figure}[tb!]
    \centering
    \includegraphics[width=0.98\linewidth]{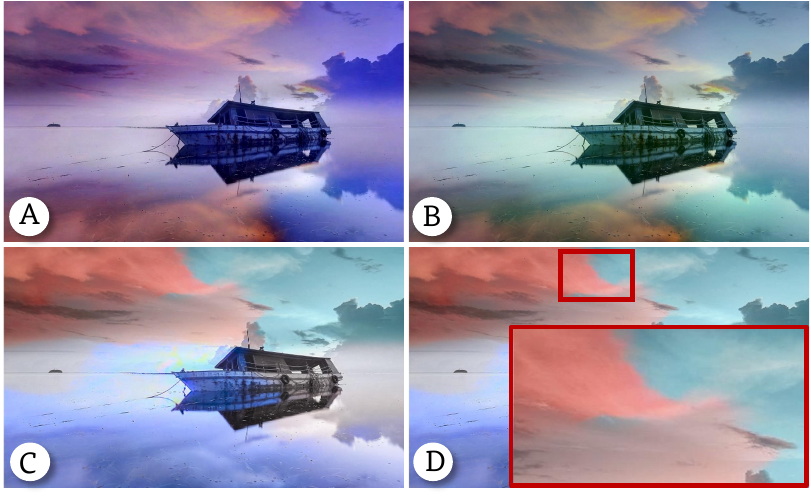}
    \vspace{-4mm}
    \caption{Comparison of designs using 1D+ format (A and B) and 2D format (C). Image D is a zoomed-in view of C, highlighting the difficulties participants experienced with the 2D palette in the sky area, which lacks distinct regions.} 
    \label{fig:task1_1_landscape}
    \Description{The figure compares designs created using different palette formats. Designs A and B utilize the 1D+ format, showcasing two distinct outcomes that demonstrate the versatility of this palette style. Design C uses the 2D format, offering a broader, more complex color manipulation that differs significantly from the 1D+ results. Image D provides a close-up view of a specific area in design C, specifically the sky region, to highlight the challenges users faced due to the absence of clear, distinct regions for color application in the 2D palette format.}
\end{figure}

\begin{figure}[tb!]
    \centering
    \includegraphics[width=0.98\linewidth]{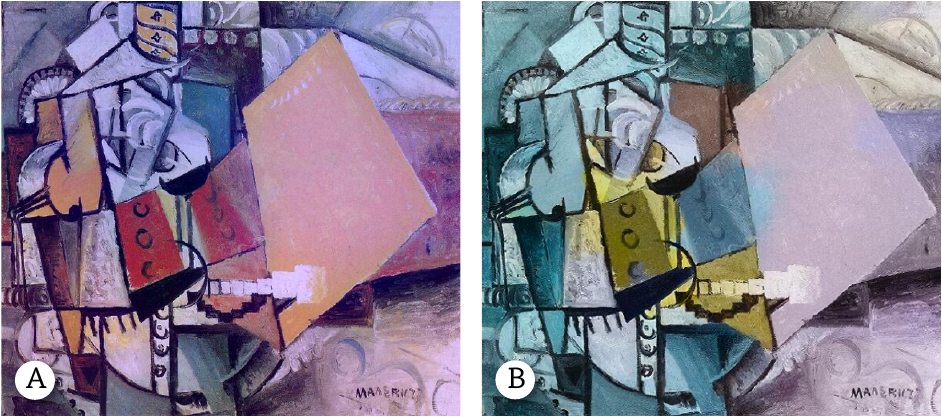}
    \vspace{-3mm}
    \caption{Comparison of designs with 1D+ format (A) and 2D format (B). Participants struggled to create the desired contrast in A but achieved it in B using the 2D palette to apply different colors to distinct regions.} 
    \label{fig:task1_1_artwork}
    \Description{The figure presents a side-by-side comparison of designs created with two palette formats. Design A, using the 1D+ format, shows an attempt that fell short of achieving the desired contrast, indicating the limitations participants faced with this simpler palette. Design B, crafted with the 2D format, successfully demonstrates the application of various colors to distinct regions, achieving the contrast that was lacking in Design A. This contrast highlights the effectiveness of the 2D palette in allowing for more nuanced and region-specific color applications.}
\end{figure}

\textbf{Expressiveness.}
The participants agreed that presenting more information and enabling more diverse adjustments in the palette enhanced the expressiveness, but the extent depended on the type and characteristics of the images being worked on. 
We observed that participants worked better with the 1D proportional palette for images without clear regions while preferred the 2D palette for images with clearly defined regions.
For instance, when working with the landscape image in Task 1, P8 and P12 skillfully utilized the proportional 1D palette to create the color blending and gradients effect for the sky.
They achieved this by initially allocating a greater proportion of red and subsequently reducing its saturation and proportion in other blocks while making subtle changes to the hue, resulting in a sunset effect that was richly layered.
P8 commented \qt{the proportion control makes it super easy to make layered and smoothly blended color effects.}
But in the 2D palette, it becomes more difficult because \pqt{you can't just slap colors on the block by block. It is tricky to blend the colors together in a more natural and random way to create a sky and clouds that look like the real}{P8}. 
We show P8's attempt with 1D+ and 2D palettes in Figure~\ref{fig:task1_1_landscape}.
In such cases, participants \pqt{don't want to be bothered by how to place them}{P12}.
Instead, the spatial 2D palette offers participants greater expressiveness in images with clearer segmentation, such as the abstract artwork in Task 1.
Many participants were creative in using colors at different spots to make some shapes pop up: \pqt{I love that I can make the left-side all blue and right-side purple and then make the center stand out with bright, vibrant yellow to create a cool contrast effect}{P1}, which is hard to achieve in 1D formats, as shown in Figure~\ref{fig:task1_1_artwork}.

\textbf{Color granularity.}
It was observed that different color granularity levels were desired by participants. 
For 1D uniform and proportional formats, two typical user behaviors were identified. 
A majority of participants (8/12) preferred a smaller range of colors (4-6) to work with, as \pqt{it is more efficient to create the overall tone [they desire]}{P12}. 
However, some users favored a larger range of colors (typically 8-12), as this \pqt{allows [them] to fine-tune the visually important but proportionally tiny colors}{P5}. 
Nevertheless, when working with many colors in 1D proportional formats, there is a scalability issue: \pqt{it becomes difficult to manipulate colors with intricate proportions due to the narrow blocks and closely placed edges}{P9}.

On the contrary, all participants preferred a larger range of colors (>8) in the 2D palette format, as they expected to manipulate more fine-grained details. 
Furthermore, participants suggested a hierarchical block system within the 2D palette, where they can zoom in on each block to achieve more detailed control within a single block. 
For instance, \pqt{a large block could represent that is the sky, but I wish I had the flexibility to zoom in and manipulate the detailed color [transitions for the sky] within that block would be nice}{P10}.

\subsubsection{Comparison of workflows between Photoshop and \sysname{} (\textbf{RQ3})}
\label{sec:finding_task2}
Overall, we identified three aspects in terms of how users perform color design tasks using the abstraction-driven approach compared to the traditional concrete-driven approach.

\begin{figure*}[htbp]
    \centering
    \includegraphics[width=\linewidth]{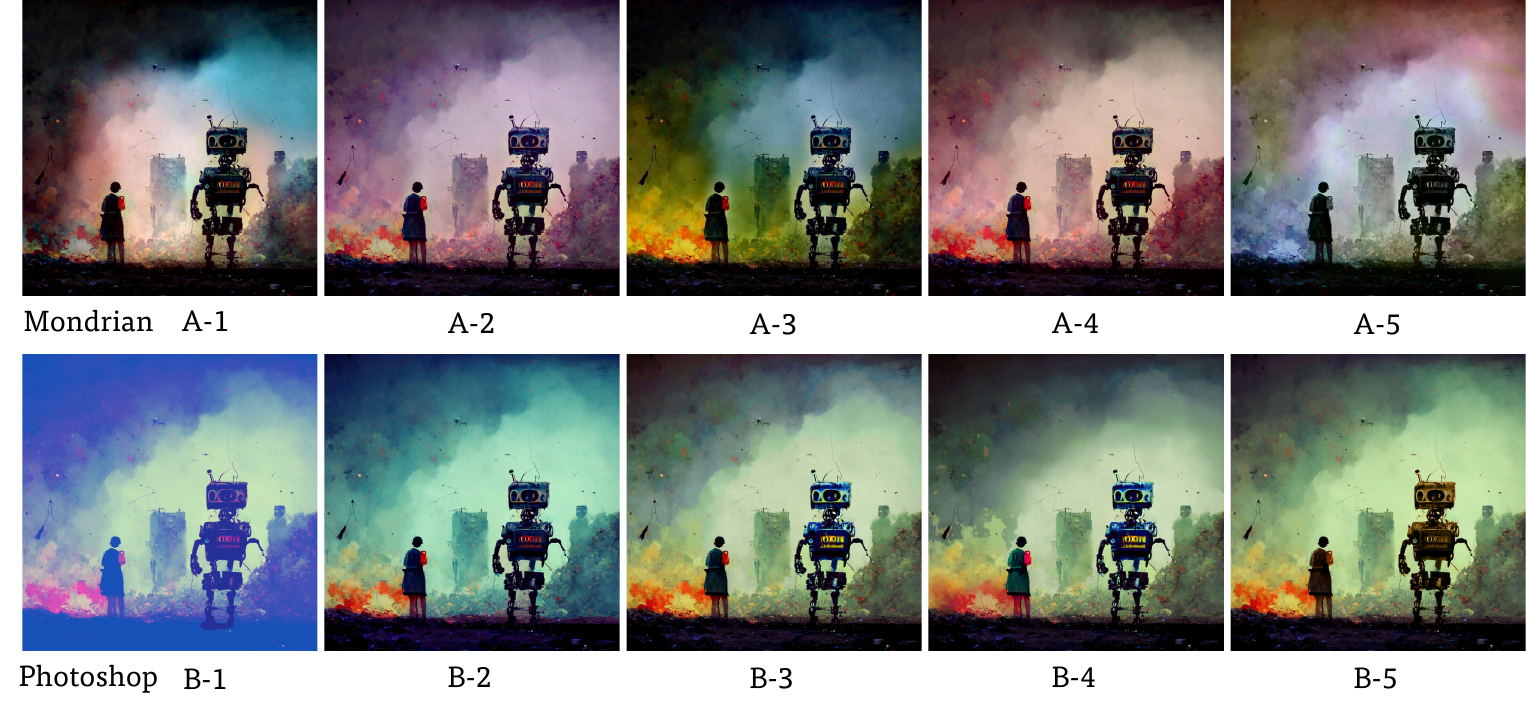}
    \vspace{-9mm}
    \caption{Participants generated designs during Task 2 in \sysname{} (first row) and Photoshop (second row). } 
    \label{fig:task2_1}
    \Description{The figure showcases two rows of designs generated by participants, corresponding to Task 2 in the study comparing Mondrian and Photoshop. The first row displays designs created with Mondrian, illustrating its capabilities through various creative outputs. The second row features designs produced in Photoshop that look more similar, offering a comparative look at how the same task is approached using a different tool.}
\end{figure*}

\begin{figure}[htb!]
    \centering
    \includegraphics[width=0.95\linewidth]{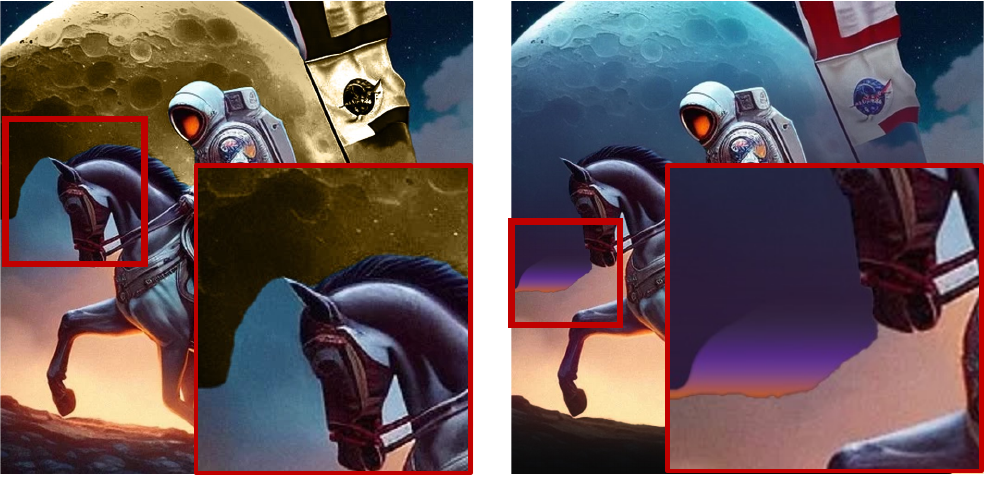}
    \vspace{-3mm}
    \caption{Two examples where participants struggled in Photoshop to refine edges.} 
    \label{fig:task2_2_edge}
    \Description{The figure presents two examples highlighting challenges faced by participants in Photoshop, specifically focusing on refining edges. The images showcase areas where the attempt to achieve precise edge definition did not meet expectations.}
\end{figure}

\textbf{Creative exploration and experimentation.}
Overall, participants found \sysname{} useful for exploring different design possibilities due to its simplicity, color-focused abstraction, and flexibility provided by switching between the three palettes. 
Compared to Photoshop, \sysname{} encourages a more flexible workflow and receptive mindset in color design exploration, ultimately leading to a more diverse range of design possibilities.

Regarding workflows, participants with Photoshop employed either a \emph{bottom-up} or a \emph{top-down} manner.
Those who adopted the \emph{bottom-up} strategy often focused on minute details of the image, spending considerable amounts of time refining specific elements one by one. 
For instance, P4, P7, and P10 dedicated a significant portion of their time to perfecting the edge of the moon (as shown in Figure~\ref{fig:task2_2_edge}): \pqt{I want to ensure the boundary appeared natural}{P10}. 
\rev{Such meticulous attention to the boundary detail shifts participants' focus away from colors, as P7 noted \qt{colors become the less.}
Consequently, the final design often appears only marginally different from the original image in terms of color.
}
Others who adopted the \emph{top-down} way would typically first modify the global hue and/or saturation then select some objects in the foreground to modify their colors. 
This approach may limit the potential for creating a richer and more diverse design, as P8 commented that \qt{though the colors are different, the style is quite fixed to me, perhaps because the hue and the saturation are linearly changed for the background.}
However, participants with \sysname{} did not follow a fixed workflow and instead had the flexibility to switch between the different palettes and freely experiment with global and local changes. 
Many participants found this approach to be more intuitive and allowed for more creative exploration. 
For example, some participants enjoyed \pqt{keep switching between the 1D proportional palette and the 2D one to work on the overall color combination and assign colors for the specific region back and forth}{P7}.

Regarding the mindset, participants with Photoshop tended to approach their design process with a preconceived plan in mind, stating that they \pqt{have everything well-planned with a roughly clear desired effect before [they] start to do}{P10}. 
Such preconceived plans, while they can lead to more targeted and specific designs, may limit the exploration of creative possibilities. 
\pqt{When having a clear image in mind, I just work to achieve that one, and then I will follow this style to further modify for the other [version]}{P12}. 
This may result in designs that are predictable or overly similar to previous works.
In contrast, participants with \sysname{} demonstrated a more open-minded approach to design, with one participant stating that they \pqt{didn't have a specific imagined [outcome] in [their] head and [were] curious to see what it would be in \sysname{}}{P10}, and \pqt{map the [generated] result back to the palette [they had specified] to make sense of it}{P1}.
Such openness allows for more experimental exploration because \pqt{some unexpected [generated] effect will inspire me to pursue a very different style}{P3}.

\textbf{Flexibility and control precision.}
Photoshop is expected to offer more flexibility and control precision than \sysname{}, because the abstraction-driven approach of \sysname{} inherently sacrifices some controlled freedom for creative support. However, our observations indicate that participants' perceived control precision and flexibility in both systems were highly dependent on their proficiency with Photoshop.

Participants who were more familiar with Photoshop were more comfortable using different tools in combination and thus could more precisely control the desired effect. 
For example, in Task 2, all participants attempted to recolor the background of the robot image in Photoshop, but only a few (2/6) succeeded. 
P7, who had excellent Photoshop skills, used the ``magic wand tool'' to isolate the sky into several layers and replace their color separately, then used the ``blur tool'' to blend each layer naturally. 
\pqt{I particularly like the different tools tailored to different kinds of parts in the image, I can fully express what I want just as I am a painter using different materials}{P7}.
However, other participants failed to experience such flexibility and precise control; they failed to find a proper tool for separating layers in the sky and eventually gave up. 
They either use ``adjustment layers - hue/saturation'' to globally adjust the entire image color or only modify the robot object and the person, both far away from their initial desired effect.
P2 commented \qt{I believe Photoshop is more powerful than this, but I am just kind of overwhelmed, so always stick to using this one [globally adjust hue and saturation].}

In contrast, those who are inexperienced with Photoshop found \sysname{} \qt{more approachable and expressive,} benefiting from the flexibility and control provided by the unique features of three palette formats. 
For example, P3 mentioned that they could \qt{easily create colorful layers for the sky using the 2D palette, and the layered effect is well-represented,} while P4 appreciated the \qt{flexibility of the 1D proportional palette for global color combinations, which is not possible in Photoshop where all colors move towards the same hue if I don't want to do selections.} These findings suggest that \sysname{} may serve as a more intuitive and user-friendly tool for color design exploration, particularly for novice designers.

\section{Survey Study} \label{sec:survey-study}

While the user study offered valuable insights into users' experience with the palettes, how these palette formats differ quantitatively in terms of people's perception of color composition remains unknown. 
To answer \textbf{RQ4}, we sought to conduct a follow-up survey study investigating the effect of palettes more systematically. 
Particularly, we aimed to understand how individuals perceive and respond to varying palette formats regarding aesthetic quality and affective feelings.

\subsection{Study Software and Dataset}
\begin{figure*}[tb]
    \centering
    \includegraphics[width=0.9\linewidth]{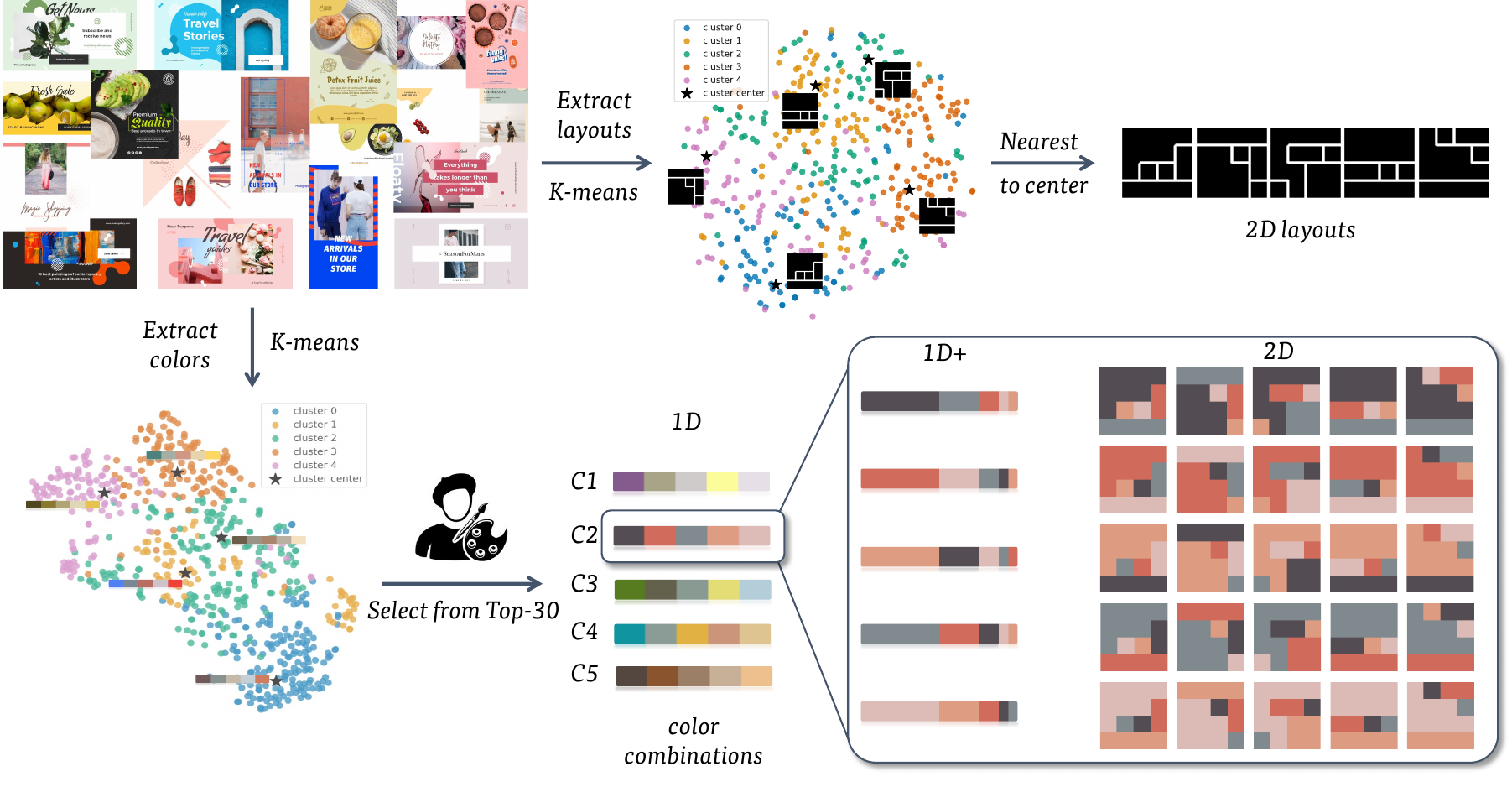}
    \vspace{-5mm}
    \caption{The process of generating test samples for the survey study and the final five color combinations used for testing, labeled C1 through C5. The final samples used in the study are showcased for color combination C2, where we display the complete 1D+ and 2D palettes. }
    \label{fig:survey_study_process}
    \Description{The figure outlines the methodology for creating test samples for a survey study, culminating in five distinct color combinations labeled C1 through C5. It also specifically showcases the final samples used in the study for color combination C2. This detailed view includes both the complete 1D, 1D+, and 2D palettes associated with C2, providing a clear visual representation of the color schemes participants evaluated.}
\end{figure*}


\begin{figure}[tb]
    \centering
    \includegraphics[width=0.95\linewidth]{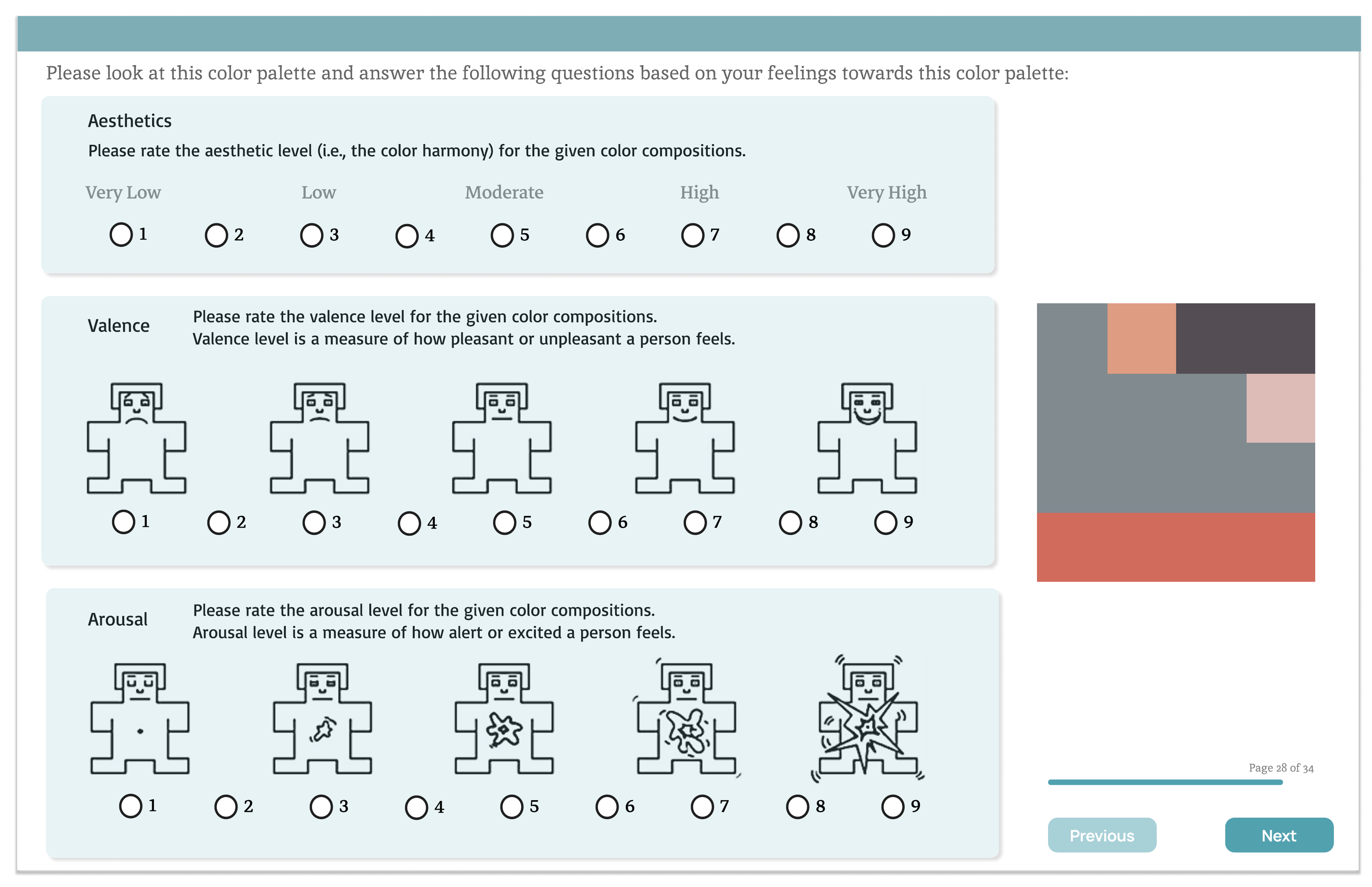}
    \vspace{-4mm}
    \caption{The interface of the survey application, consisting of three questions for participants to rate regarding aesthetics, valence, and arousal. The color palette is displayed on the side. Participants rate one palette at a time, then click the ``Next'' to rate the next.}
    \label{fig:survey_study_interface}
    \Description{The figure depicts the interface of the survey application designed for participants to evaluate color palettes. It includes three questions regarding aesthetics, valence, and arousal. Besides the questions, the color palette under evaluation is prominently displayed on the right side, allowing participants to reference it while providing their ratings. Participants are instructed to rate one palette at a time and then proceed to the next by clicking the ``Next'' button.}
\end{figure}

In the survey, we considered the three color palettes used in \sysname{}, 1D, 1D+, and 2D.
We developed a web-based application, shown in Figure~\ref{fig:survey_study_interface}, that displayed different color palettes subsequently and asked respondents to rate the levels of harmony (aesthetics) as well as valence and arousal (affectiveness) \cite{posner2005circumplex} of each color palette. 
Each color palette presented five different colors. 
We chose five colors as this is the most commonly found configuration on existing platforms, such as Adobe Color.
In the case of 1D+, each color was also displayed with its proportion; and with 2D, with the same color proportions, their spatial placement was presented as well.  


To ensure the color combinations, proportions, and spatial placements used in the study were representative, we adopted the following approach to extract color palettes from a graphic design dataset \cite{destijl2023shi}.
To decide on which colors to use, we extracted each design's uniform 1D color palette, consisting of five dominant colors. 
Next, we performed k-means clustering with five clusters and selected the top 30 palettes nearest to the center for each cluster as the palette candidates. 
Since color perception is highly subjective and the palette nearest to the center may not necessarily be the most representative, we sought the help of a design expert to select one palette from the candidates for each cluster.
This individual had undergone systematic color theory training and had more than eight years of experience in graphic design.
The selected palettes decided the color combinations, and we used a random order for presenting the color blocks in the 1D palettes.
This step produced all the 1D color palettes used in the survey. 

To decide the color proportions, we employed the same approach described in Sec~\ref{sec:palette_extraction} to extract five-color 2D palettes from the designs.
We then computed the 2D areas of the blocks with the same color in these 2D palettes and sorted them. 
By manually checking the areas, we obtained the representative proportion which was approximately 8:4:2:1:1. 
Using this proportion, we were able to augment a 1D palette in the previous step to the 1D+ format.
Regarding the order of the colors in 1D+ palettes, we sorted the colors by their proportions from the largest to the smallest. 

Lastly, to decide the color placement for 2D palettes, we standardized the extracted 2D palettes in the previous step by replacing the colors with black blocks and white edges, so only layout information remained.  
Next, we performed k-means clustering on these spatial layouts with five clusters, and chose the ones nearest to the cluster center with the color proportion 8:4:2:1:1. 
In this way, we could ensure the 2D palettes had the same proportion as the 1D+ ones.
Using the selected five layouts, we could augment a 1D+ palette in the previous format to five 2D ones.



\subsection{Participants and Design}
We recruited 100 participants from the Prolific platform \cite{prolific}. 
We pre-screened the participants to ensure that the selected individuals did not have any color blindness or difficulties in distinguishing between certain colors. 
Participants were compensated at a rate of \rev{CAD\$14} per hour for their time and effort.

We utilized a mixed design with both between-subjects and within-subjects factors. 
Our between-subjects factor was color combination, where we investigated five different combinations, C1 -- C5 in Figure~\ref{fig:survey_study_process}.
Therefore, each color combination condition had 20 participants, who were randomly assigned upon recruitment.
For each color combination, our within-subjects factors included the color palette format, as well as the color proportion and spatial placement that 1D+ and 2D exhibited. 
A total of 31 conditions for the color palettes were presented to each participant (Figure~\ref{fig:survey_study_process}), including one for the 1D format, five for the 1D+ format, and 25 for the 2D format (5 proportions $\times$ 5 placements). 
For each condition, we asked participants to rate the harmony, arousal, and valence of the color palette on a 9-point Likert scale (1=very low; 9=very high).
The order of the conditions was randomized to address the potential ordering effect.

\subsection{Results}

\begin{table*}[tb]
\caption{Statistical results of the Kruskal-Wallis tests and adjusted \textit{p}-value of pairwise comparisons. Orange cells indicate significant differences ($p <0.05$) and blue cells indicate non-significant differences ($p>= 0.05$).
\vspace{-3mm}
}
\label{tab:survey_results}
\small
\begin{tabular}{l|c|c|c|c|c}
\toprule
\textbf{Metric} & \multicolumn{1}{l|}{\textbf{Color Comb.}} & \multicolumn{1}{l|}{\textbf{Across 1D, 1D+, \& 2D}} & \multicolumn{1}{l|}{\textbf{1D vs. 1D+}} & \multicolumn{1}{l|}{\textbf{1D vs. 2D}} & \multicolumn{1}{l}{\textbf{1D+ vs. 2D}} \\
\toprule
\multirow{6}{*}{\textbf{Harmony}} & All & $H(2) = 57.582$, $p < 0.001$ & $p<$ 0.001 & $p<$ 0.001 & $p<$ 0.001 \\
 & C1 & $H(2) = 12.567$, $p < 0.01$ & $p=$ 0.205 & $p<$ 0.05 & $p<$ 0.05 \\
 & C2 & $H(2) = 8.127$, $p < 0.05$ & $p<$ 0.05 & $p<$ 0.05 & $p=$ 0.829 \\
 & C3 & $H(2) = 1.089$, $p = 0.580$ & - & - & - \\
 & C4 & $H(2) = 17.791$, $p < 0.001$ & $p<$ 0.01 & $p<$ 0.001 & $p=$ 0.119 \\
 & C5 & $H(2) = 22.904$, $p < 0.001$ & $p<$ 0.01 & $p<$ 0.001 & $p<$ 0.05 \\
\midrule
\multirow{6}{*}{\textbf{Valence}} & All & $H(2) = 49.392$, $p < 0.001$ & $p<$ 0.001 & $p<$ 0.001 & $p<$ 0.01 \\
 & C1 & $H(2) = 10.000$, $p < 0.01$ & $p=$ 0.313 & $p<$ 0.05 & $p=$ 0.069 \\
 & C2 & $H(2) = 12.436$, $p < 0.01$ & $p<$ 0.05 & $p<$ 0.01 & $p=$ 0.450 \\
 & C3 & $H(2) = 4.473$, $p = 0.107$ & - & - & - \\
 & C4 & $H(2) = 13.067$, $p < 0.01$ & $p<$ 0.05 & $p<$ 0.01 & $p=$ 0.198 \\
 & C5 & $H(2) = 10.204$, $p < 0.01$ & $p<$ 0.01 & $p<$ 0.01 & $p=$ 0.851 \\
\midrule
\multirow{6}{*}{\textbf{Arousal}} & All & $H(2) = 16.29$, $p < 0.001$ & $p<$ 0.05 & $p<$ 0.001 & $p=$ 0.065 \\
 & C1 & $H(2) = 10.741$, $p < 0.01$ & $p=$ 0.155 & $p<$ 0.05 & $p=$ 0.110 \\
 & C2 & $H(2) = 3.337$, $p = 0.189$ & - & - & - \\
 & C3 & $H(2) = 1.437$, $p = 0.487$ & - & - & - \\
 & C4 & $H(2) = 7.002$, $p < 0.05$ & $p=$ 0.066 & $p<$ 0.05 & $p=$ 0.759 \\
 & C5 & $H(2) = 1.116$, $p = 0.572$ & - & - & - \\
\bottomrule
\end{tabular}
\end{table*}

As shown in Table~\ref{tab:survey_results}, Kruskal-Wallis tests across all the trials revealed significant differences in all three metrics (\ie, harmony, valence, and arousal) for palette format, which indicates that color palettes indeed play an important role in users' perception of color compositions. 
Another set of Kruskal-Wallis tests revealed significant differences among the five combinations for all perceptual metrics as well: harmony ($H(4)=76.774, p<.001$), valence ($H(4)=58.233, p<.001$), and arousal ($H(4)=56.033, p<.001$).
This indicates that working with different colors can affect how users perceive them using different palette formats, which led us to investigate the effects of color palettes within each color combination.

The results are summarized in Table~\ref{tab:survey_results}.
Within each color combination, significant differences were observed in harmony and valence for all combinations except for C3.
In contrast, for arousal, significant differences were only observed in C1 and C4. 
These findings suggest that the choice of color palette format may have a more pronounced impact on the harmony and valence dimensions, whereas its effect on arousal appears to be more selective and confined to specific color combinations. 

W also conducted post-hoc pairwise comparisons using the Tukey HSD test to pinpoint the specific differences among the three palette formats while applying a multi-testing p-value correction (Table~\ref{tab:survey_results}).
When testing on all the trials, significant differences were identified across all three dimensions in pairwise comparisons, with the exception of the 1D+ vs. 2D comparison for arousal.
Within each color combination, pairwise comparisons indicated that 1D vs. 2D consistently yielded significant results across all three metrics, which suggests that presenting a given color combination in either a 1D or 2D format can influence human perceptions.
For 1D vs. 1D+, significant differences were observed for most combinations on harmony and valence, while no significant differences were noted with respect to arousal. 
In contrast, 1D+ vs. 2D consistently displayed no significant differences across all color combinations for valence and arousal, although significant differences were observed occasionally for harmony.

We summarize the key takeaways as follows: 
\begin{itemize}
    \item Color combinations significantly influence perceptual metrics including harmony, valence, and arousal. 
    \item Palette formats affect perceptions, with the impact level varying across specific color combinations. 
    \item The distinction between 1D and 2D palettes significantly influences users' perception of color compositions, whereas the influence of 1D+ vs. 2D is less consistent and lacks substantial evidence supporting its significance. 
    This suggests that the combined effect of proportion and spatial placement variations in palettes has a more substantial impact than that of either factor individually.
\end{itemize}

\section{Discussion}
\rev{
In this section, we discuss the design implications of our findings and future directions to improve the current design of \sysname{} to better support design ideation, as well as the limitations of our tool and studies.
}

\subsection{Future Design Considerations and Directions} \label{sec:design-considerations}
Our study has provided the following design considerations that could shed light on the future design of similar tools.

\textbf{Exploring the full spectrum of abstract to concrete manipulation.}
Our user study revealed that approaching colorization tasks in a progressive manner, moving from abstract to concrete controls along a spectrum, benefits exploring creative possibilities and reducing task workloads. 
Envisioning this spectrum, we have \sysname{} at one end providing high-level abstraction, while at the other end, tools like Photoshop deliver detailed and precise control. 
Color palette formats such as 1D+ and 2D act as intermediate stepping stones within this progression.
These different palette formats play a crucial role in offering users various levels of control and enabling them to focus on different aspects, such as overall tone adjustments with the 1D+ format or local refinements with the 2D format.
Though \sysname{} provides flexibility, this control spectrum could be more densely populated.
To further enhance user control in future interactive tools like \sysname{}, we suggest integrating hierarchical structures into the existing palettes.
The current color blocks within the palettes could be treated as the primary level, and users can access secondary or even tertiary levels for increased precision.
For example, in the 1D+ palette, entering the secondary level would allow users to specify how a particular color can be subdivided into additional colors with subtle differences in hue and saturation, resulting in a more nuanced effect. 
Similarly, in the 2D palette, accessing the secondary level would enable users to focus on smaller regions or distinct objects within the image, allowing for more intricate manipulation.
Incorporating hierarchical structures into interactive color design tools will provide users with a more dynamic range of control options, effectively bridging the gap between abstract and concrete manipulation in the spectrum and leading to more comprehensive and efficient design processes.

\rev{\textbf{Investigating advanced color palette visualizations and adapted recolorization models.}}
\rev{
In our study, we evaluated three distinct color palette formats, each offering specific levels of control and facing unique limitations. 
For instance, the 2D palette fails in images with soft gradients, a limitation caused by the corresponding colorization method \cite{zhang2017real}. 
Other unexplored palette formats in the literature may also have their advantages while presenting their own drawbacks, such as Color Triads \cite{shugrina2020nonlinear} which offers effective color gradient interpolation but is not suitable for images having more than three color themes due to their triangular structure.
This highlights the need for designing a comprehensive set of palette formats with greater descriptive power to cover different use cases. 
Current palettes, such as 1D+ and 2D, provide insights into color proportions and spatial placements. 
Their combination, as demonstrated in Task 2 of our user study, effectively covers a wide range of images and exploratory purposes.
However, further exploration is needed to investigate additional characteristics, such as identifying individual objects or distinguishing foreground from background in images. 
It is important but challenging to effectively integrate the identified characteristics into both palette designs and recolorization models, for instance, how to accurately reflect those specified characteristics on the recolored images while ensuring high image quality and reducing the occurrence of artifacts. 
For future research in abstraction-driven approaches, we suggest exploring optimal combinations of visual abstractions and image features, and pairing them with tailored recolorization models to accommodate broader use cases.
}

\textbf{Establishing an explicit mapping between the palette and the image.}
In our current interface design, the image and the palette editing panels are placed side by side. 
However, for the 2D palette, some participants suggested directly overlaying the 2D palette onto the image. 
This idea stems from the fact that the 2D palette is designed to reflect color placements, prompting users to mentally associate each color block in the 2D palette with a specific region in the image. 
Although all participants agreed that this mental mapping is intuitive, it can sometimes be misaligned with the actual image, particularly when the width and height of the image are unequal, whereas the current 2D palette consistently maintains a square shape.
Users are forced to mentally ``stretch'' the palette to align it with the image, which can result in mismatches. 
By overlaying the reshaped 2D palette onto the image, users may gain a more accurate understanding of the specific regions they are working with. 
Another potential improvement to address such mismatch issues is to dynamically link the palette and the image---when users manipulate the palette, the corresponding regions of the image could be highlighted. 
Both of these enhancements could potentially improve users' overall design experience.

\rev{
\textbf{Providing flexibility under a unified manipulation logic.}
Our study demonstrates that \sysname{}, as a tailored tool for abstraction-driven image colorization, effectively facilitated rapid color design exploration due to its simplicity. 
In contrast, a general-purpose tool like Photoshop, while offering a greater set of operations, presents a steeper learning curve, better suited for detailed design refinement and other manipulations rather than color design ideation.
However, it is not surprising that a fit-for-purpose tool excels at the specific tasks of interest more than a general-purpose tool.
Further, our participants valued the degree of flexibility and the variety of operations, even in the ideation stage.
Therefore, there exists a trade-off between simplicity (\ie, focusing on specific tasks) and flexibility (\ie, supporting various functionalities).
We argue that the conflict between simplicity and flexibility can be mitigated by carefully designing the features within a unified manipulation logic.
For instance, in \sysname{}, different palette formats and manipulation approaches are all aligned with the same interactive palette logic, offering flexibility without significantly adding a user's cognitive load. 
On the other hand, a wide array of functionalities integrated in a tool may be based on different manipulation logics, while each optimized for specific tasks. 
For example, in Photoshop, users can recolorize an image by selection-based approach and layer-overlayed approach, which are two different logics to remember. 
Therefore, we posit that the learning curve is influenced by not only the number of features but also how the varieties of manipulation logics. 
Thus, designing tools for rapid exploration requires a careful consideration of how much flexibility and control can be offered within a single, coherent manipulation logic, which echos Nielsen's usability heuristic of ``recognition rather than recall'' \cite{nielsen1994enhancing}. 
Future research could design rigorous comparison studies to understand how the number of manipulation logics and the breadth of features impact users' learning difficulty, exploration, and controllability. 
}

\subsection{Limitations}
Our tool and studies still have a few limitations that we should acknowledge. 
First, in current \sysname{}, we empirically decided on a color quantity range from 4 to 12. 
However, according to our study, participants preferred different color quantity ranges for different palette formats, particularly for the 2D palette, where they tended to require more colors. 
Future research could explore the optimal color quantity range for each palette format, taking into account user preferences and task requirements.
Second, participants expressed appreciation for Photoshop's real-time preview feature, which is currently not available in \sysname{} due to the inference time of the colorization models. Nevertheless, incorporating real-time preview is essential for providing users with immediate feedback on their design choices and enhancing the overall design experience. 
Third, the participants in our survey study were mostly unfamiliar with color theory and lacked design training. 
Consequently, the conclusions drawn from the survey study can primarily be applied to novice users. 
It remains unclear whether the observed differences in perception towards color compositions would hold for a wider group of users, including expert designers. 
Future research should consider conducting similar studies to determine whether the findings are generalizable across varied expertise levels.

\section{Conclusion}
This paper presents an extensive exploration of discrete color palettes as interactive abstractions for color design ideation.
Based on the user-centered design process, we introduced \sysname{}, a system enabled \emph{abstraction-driven} color manipulation to boost color design exploration.
\sysname{} implements three interactive color palettes integrated with AI colorization models, serving as a test-bed to explore how visual abstractions reshape the ideation process. 
Through an in-lab user study, we gained insights into how users interact with color palettes as interactive abstractions and the impact on their workflows and cognitive creativity compared to traditional \emph{concrete-driven} manipulation.
\rev{
The findings revealed the benefits and challenges of the three interactive color palette formats (\ie, 1D uniform palette, 1D proportional, and 2D spatial palettes), leading to the design implications on future visual abstraction design.
In addition, our research confirms the efficacy of \sysname{} in enabling users to creatively explore various design possibilities and effectively accomplish color design. 
}
With a survey study involving 100 participants, we discovered the significant role of combined proportion and spatial placement variations in color palettes on human perception of color design aesthetics and emotion associations.
The findings and insights obtained through our studies offer implications for future research in developing more intuitive, efficient, and creative tools tailored for supporting ideation in color design and other design-related fields. 
\begin{acks}
We thank all our participants for their time and valuable input. We would also like to thank our reviewers whose insightful comments have led to a great improvement of this paper. This work is supported in part by the NSERC Discovery Grant and a gift fund from Adobe Systems, Inc.
\end{acks}

\bibliographystyle{ACM-Reference-Format}
\bibliography{references.bib}

\end{document}